\documentclass[%
reprint,
superscriptaddress,
 amsmath,amssymb,
 aps,
 prb,
]{revtex4-1}

\usepackage{graphicx}
\usepackage{dcolumn}
\usepackage{bm}
\usepackage{epstopdf}
\usepackage{subfigure}
\usepackage{underscore}
\usepackage{color}
\usepackage{ulem}
\usepackage{animate}

\begin{document}

\preprint{}

\title{Stacking-Dependent Spatial Charge Separation in Graphitic Carbonic Nitride layers}

\author{Chen Jiao}
\affiliation{
 State Key Laboratory of Luminescence and Applications, Changchun Institute of Optics, Fine Mechanics and Physics, Chinese Academy of Sciences, No.3888 Dongnanhu Road, Changchun 130033, People¡¯s Republic of China
}
\affiliation{
 University of Chinese Academy of Sciences, Beijing 100049, People¡¯s Republic of China
}
\author{Da Zhan}
\affiliation{
 State Key Laboratory of Luminescence and Applications, Changchun Institute of Optics, Fine Mechanics and Physics, Chinese Academy of Sciences, No.3888 Dongnanhu Road, Changchun 130033, People¡¯s Republic of China
}
\affiliation{Research Institute for Soft Matter and Biomimetics, College of Materials, Xiamen University, Xiamen, Fujian, 361005, China}
\author{Dong Han}
\affiliation{
 State Key Laboratory of Luminescence and Applications, Changchun Institute of Optics, Fine Mechanics and Physics, Chinese Academy of Sciences, No.3888 Dongnanhu Road, Changchun 130033, People¡¯s Republic of China
}
\author{XueJiao Chen}
\affiliation{
 State Key Laboratory of Luminescence and Applications, Changchun Institute of Optics, Fine Mechanics and Physics, Chinese Academy of Sciences, No.3888 Dongnanhu Road, Changchun 130033, People¡¯s Republic of China
}
\affiliation{
 University of Chinese Academy of Sciences, Beijing 100049, People¡¯s Republic of China
}
\author{Wei Chen}
\affiliation{Department of Chemistry National University of Singapore 3 Science Drive 3, 117543 Singapore}
\author{Hai Xu}
\affiliation{
 State Key Laboratory of Luminescence and Applications, Changchun Institute of Optics, Fine Mechanics and Physics, Chinese Academy of Sciences, No.3888 Dongnanhu Road, Changchun 130033, People¡¯s Republic of China
}
\author{Lei Liu}
 \email{liulei@ciomp.ac.cn}
\affiliation{
 State Key Laboratory of Luminescence and Applications, Changchun Institute of Optics, Fine Mechanics and Physics, Chinese Academy of Sciences, No.3888 Dongnanhu Road, Changchun 130033, People¡¯s Republic of China
}
\author{DeZhen Shen}
\email{shendz@ciomp.ac.cn}
\affiliation{
 State Key Laboratory of Luminescence and Applications, Changchun Institute of Optics, Fine Mechanics and Physics, Chinese Academy of Sciences, No.3888 Dongnanhu Road, Changchun 130033, People¡¯s Republic of China
}
\date{\today}

\begin{abstract}
We reveal the existence of stacking-dependent spatial(SDS) charge separation in graphitic carbonic nitride (g-C$_3$N$_4$) layers, with the density functional theory (DFT) calculations. In g-C$_3$N$_4$ bilayers, such SDS charge separation is found in particular effective for top valence bands that can drive electrons 100$\%$ away from one layer to the other. However, for bottom conduction ones, it results in little charge redistribution between layers. As spatial charge separation naturally suppresses the electron-hole recombination, that makes g-C$_3$N$_4$ layers with proper stacking much more efficient for harvesting solar energy in photovoltaic or photocatalytic applications. The SDS charge separation has been understood as a result of the inter-layer quantum entanglement from those g-C$_3$N$_4$ band electrons, whose unique chirality and phases in corner-atom-shared C$_6$N$_{10}$ units are relatively isolated and in tune only through the corner N atoms.

The SDS charge separation in g-C$_3$N$_4$ may lead to an intrinsic way, $i.e.$ without alien dopings, interfaces or electrical fields, to manipulate charge carriers in semiconducting materials. That may lead to new physics in the future optoelectronics or electronics of two-dimensional (2D) materials, such as realizing the layer-selected charge transport through the bilayer or multi-layer 2D materials.

\end{abstract}

\maketitle

\newpage

In recent years, 2D materials have brought quite some new physics unusual to the bulk ones, in particular since the isolation of graphene in 2004 \cite{Novoselov04,Mannix17,Geim07}. In a narrow sense, the emerging new 2D materials have been defined as those discrete atomically thin layers, one to several, stacked together with the weak inter-layer interaction, often as van der Waals forces\cite{Mannix17}. As without fixed inter-layer bondings, certainly these 2D materials can be assembled in various stacking styles, as their lubricant bulk phases, such as graphite, hexagonal boron nitride (h-BN), and molybdenum disulfide. Recently it was demonstrated that by certain small magic twist angle the electronic bands of a twisted graphene bilayer become flat near zero Fermi energy and exhibit insulating states at half-filling\cite{Cao1}. That stacking was found to result in the superior electronic performance of the two graphene sheets, $i.e.$ the superconductivity\cite{Cao2}.

For semiconducting 2D materials, although with weak inter-layer coupling, stacking may also play an magic role in tuning their electronic behaviors. In the case of h-BN, theoretically we showed with DFT calculations that by changing the stacking the bandgap of h-BN can be tuned from indirect to direct\cite{Liu03}. Experimentally the existence of direct-gap nature in h-BN had been confirmed later by the observation of ultraviolet lasing on a h-BN single crystal\cite{Watanabe04}. In this letter, we show besides bandgap tuning stacking can be used as a unique parameter to achieve complete band charge separation between layers in another van der Waals 2D material of g-C$_3$N$_4$.

  \begin{figure}
     \centering
          \includegraphics[width=25em]{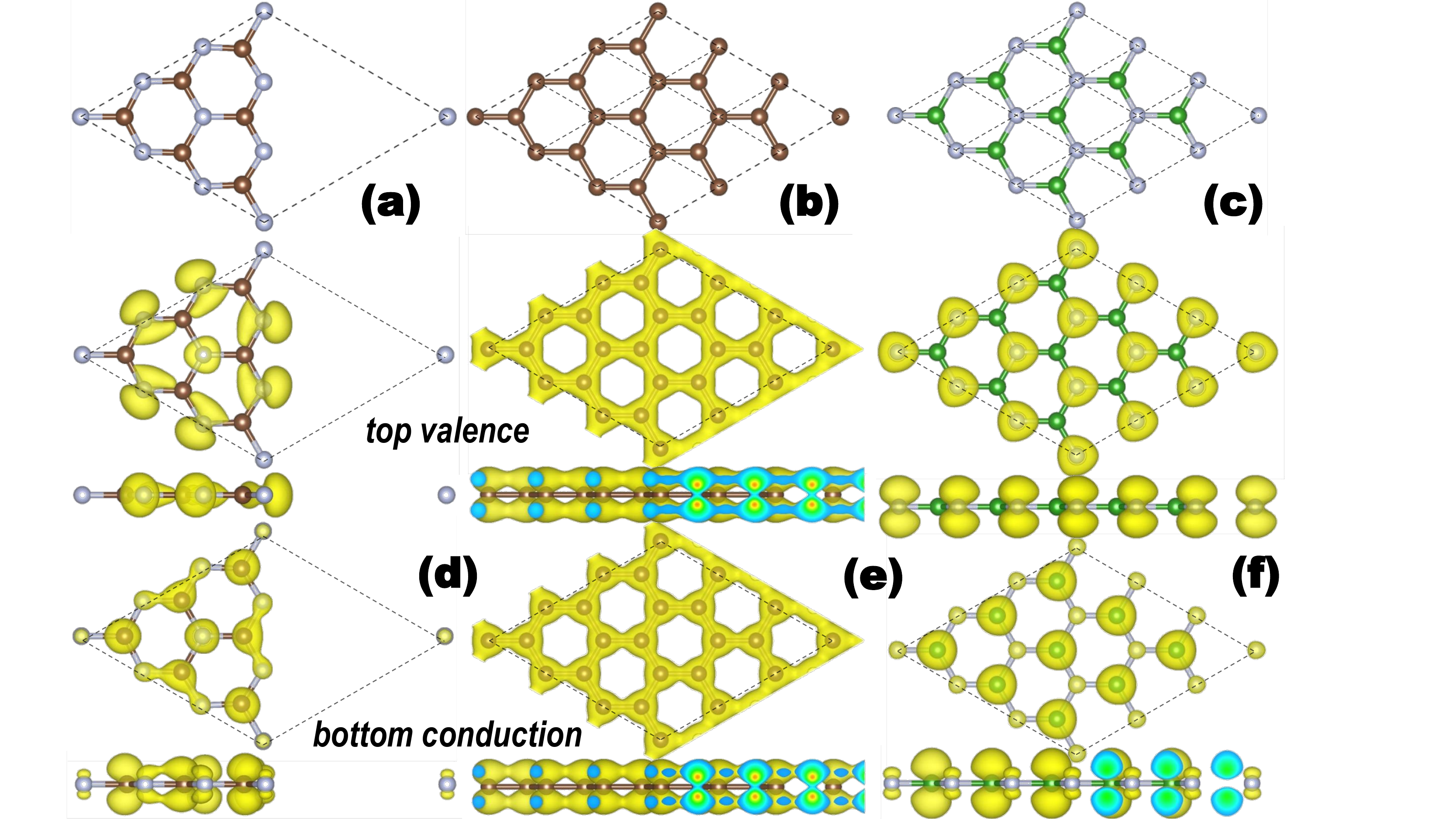}\\
          \caption{The crystal structures of single-layer g-C$_3$N$_4$ (\textbf{a}), graphene (\textbf{b}), and the single-layer h-BN (\textbf{c}); and the corresponding isosurface plots (\textbf{d}, \textbf{e}, and \textbf{f}) of their partial charge densities of top valence and bottom conduction bands; where the grey, brown, and green balls represent N, C, and B atoms, respectively.}\label{fig:fig1}
   \end{figure}

In comparison with zero-bandgap graphene and wide-bandgap h-BN, g-C$_3$N$_4$ has a medium bandgap of $\sim$2.7 eV, and possesses the appropriate positions of valence band maximum (VBM) of -1.8 eV and conduction band minimum (CBM) of 0.9 eV separately with respect to reversible hydrogen electrode\cite{Wang09}.  Moreover, through molecular structure engineering, its bandgap can be tuned to fit a wide range of visible light\cite{Liu15, Zhou15, Cui16, Kessler17} that makes g-C$_3$N$_4$ an extremely ideal photocatalytic candidate material for harvesting solar energy\cite{Zhou18}. As shown in Fig. \ref{fig:fig1}a, although half of its unit cell is empty and without atom filling,\cite{Wang09} the single-layer g-C$_3$N$_4$ has the similar honeycomb lattice as graphene (Fig. \ref{fig:fig1}b) and the single-layer h-BN (Fig. \ref{fig:fig1}c). The g-C$_3$N$_4$ single atomic layer can be taken as consisting of alternatively planar triangular C$_6$N$_8$ molecules and triangular voids. Consequently, its bonding electrons also cover only half of its lattice skeleton leaving triangle voids in relatively large proportion as shown in Fig. \ref{fig:fig1}d, unlike the even and continuous distributed bonding charges in graphene (Fig. \ref{fig:fig1}e) and the single-layer h-BN (Fig. \ref{fig:fig1}f). Thus, with stacking variation, no matter translational or rotational, electrons will sense much large amplitude of charge fluctuation between atomic layers in g-C$_3$N$_4$ than those in graphite or h-BN.

For simplicity and without loss of generality, the bilayered g-C$_3$N$_4$ is sampled in this work to investigate the electronic behaviors of g-C$_3$N$_4$ layers upon different stacking styles. Our DFT calculations are based on the Perdew-Burke-Ernzerhof (PBE) generalized gradient approximation\cite{pbe} and the projected augmented wave (PAW) method\cite{Bloch1994} with the vdw-d2 correction\cite{Grimme06, Bucko10}, as implemented by the Vienna ab initio simulation package (VASP)\cite{Kresse1994, Kresse1996, Kresse1999}. The cutoff energy for the plane-wave basis set is 520 eV and the Brillouin zone is sampled with the Monkhorst-Pack mesh of 4 $\times$ 4 $\times$1 for g-C$_3$N$_4$ layers. For the g-C$_3$N$_4$ bilayers simulated, the hexagonal lattice parameter of $a$ and the inter-layer distance are respectively set to the bulk values of 7.13 nm and 3.25 nm \cite{Wang09} , and the longitudinal cell parameter of $c$ is set to 20 nm.

\begin{figure}
      \centering
      \includegraphics[width=25em]{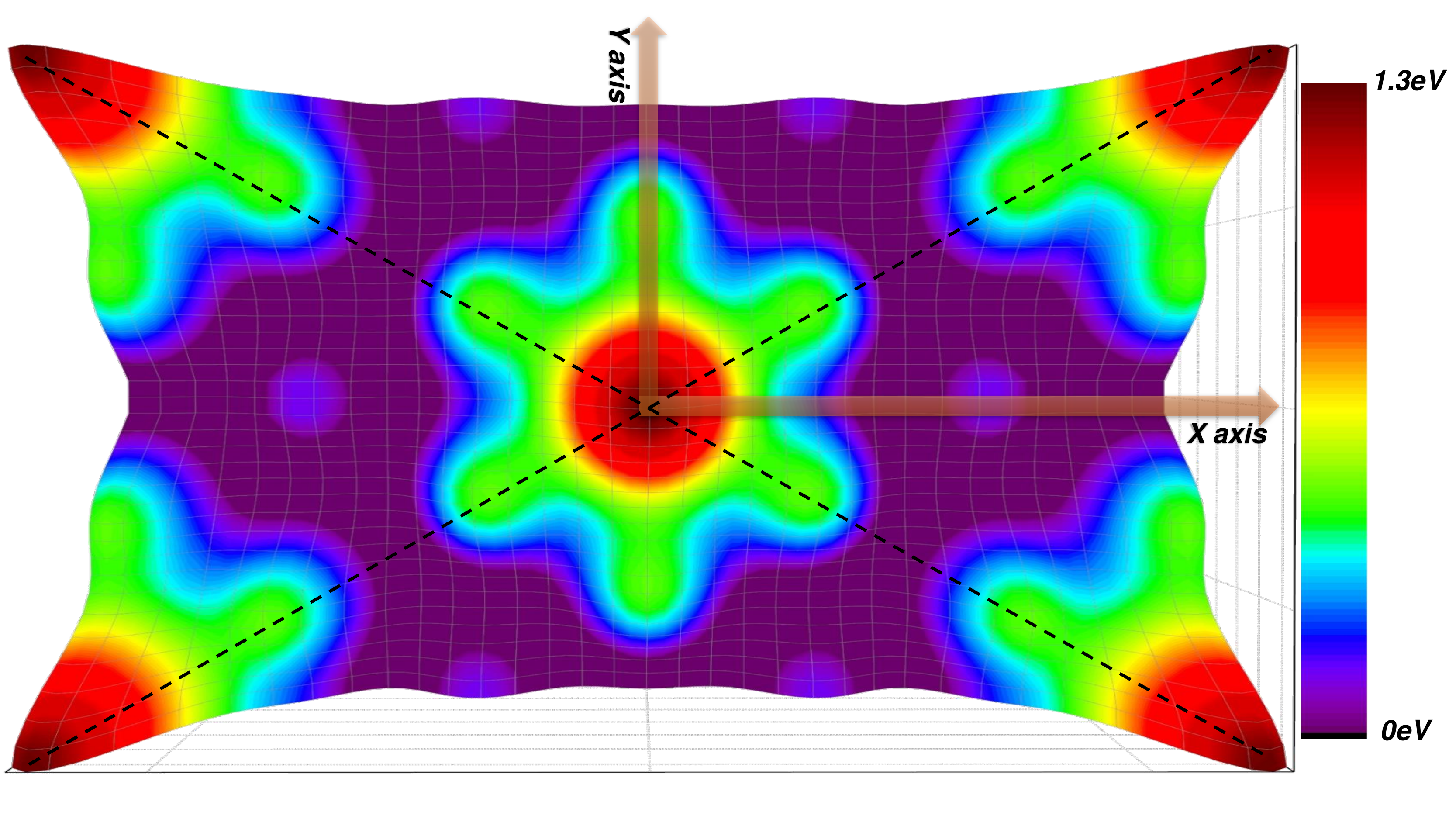}\\
      \caption{The calculated total energy of a g-C$_3$N$_4$ bilayer in dependence of their translational stacking motions from the P-6M2 stacking, where the dash lines label the basis vectors of the single-layer unit cell. }\label{fig:fig2}
\end{figure}

Fig. \ref{fig:fig2} plots the calculated total energy of the g-C$_3$N$_4$ bilayers with the translational stacking variation from the initial stacking with the symmetry group of P-6M2. The initial P-6M2 stacking is not stable as it possess the maximum energy value. And as one layer glides away from the exactly overlapping position of P-6M2 stacking, the total energy decreases by about 50 meV/atom to the energy minima in certain stacking. That indicates the inter-layer coupling of g-C$_3$N$_4$ bilayers is rather strong as expected due to their extremely inhomogeneous relative charge fluctuation with stacking. It is almost one order of magnitude bigger than the energy variation amplitude of some other van der Waals atomic layers. For examples, in bulk h-BN and graphite whose layers are coupled from both sides, the translational stacking moves can only reduce the total energy from their maxima by about 10 meV/atom\cite{Liu03}.

\begin{figure}
      \centering
      \includegraphics[width=25em]{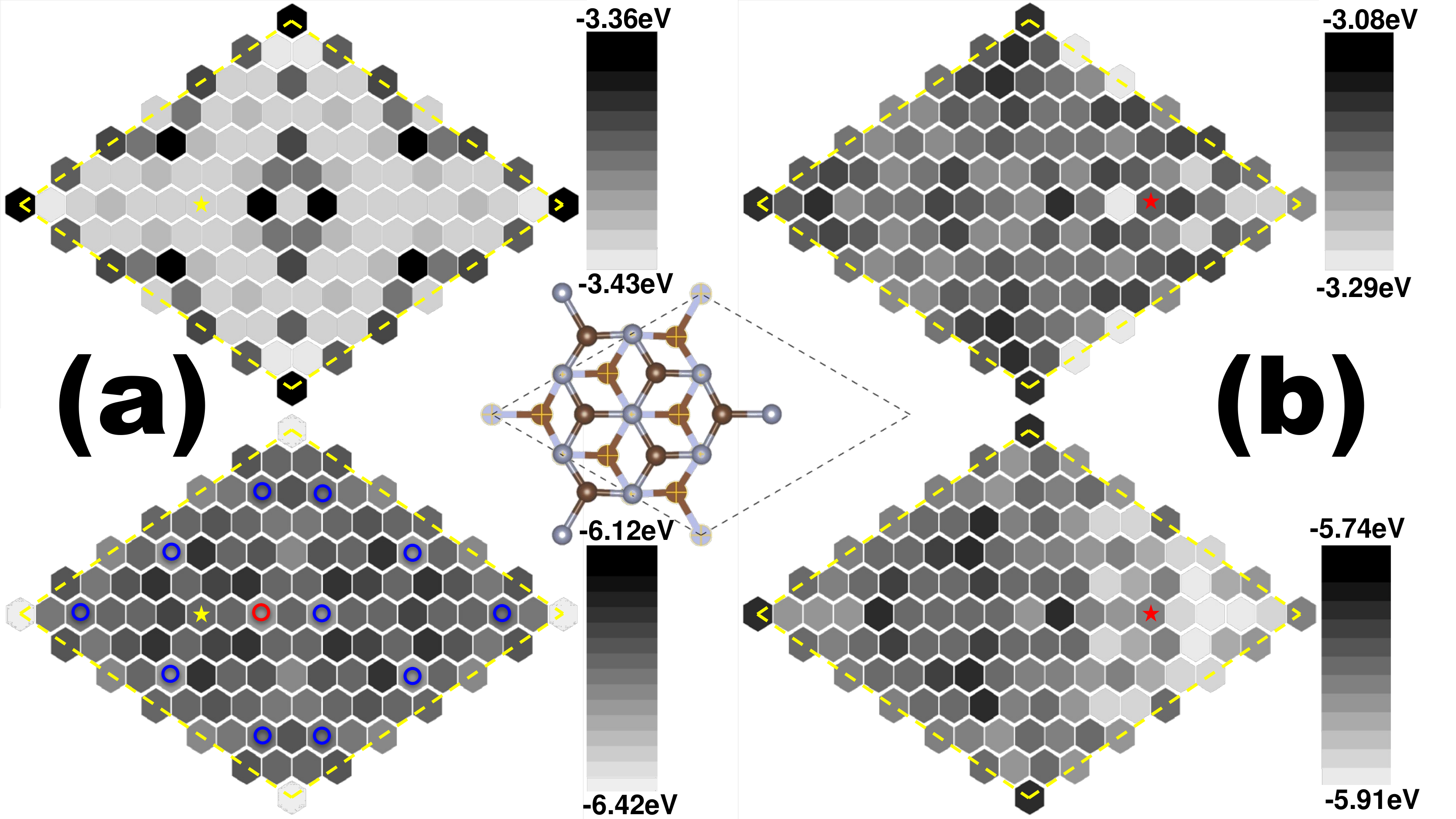}\\
      \caption{The calculated VBM and CBM of the g-C$_3$N$_4$ bilayer upon the stackings (a) translational from the P-6M2 stacking, and the blue bars plots the stackings (b) translational from the initial 60$^{\circ}$-rotation layers with the symmetry group P-3M1 as shown in the inset figure. Here the vacuum level is set to 0 eV.}\label{fig:fig3}
\end{figure}

While the inter-layer coupling of g-C$_3$N$_4$ bilayers is sensitive to stacking, it would be meaningful to examine further the stacking effect on their electronic band structures. Fig. \ref{fig:fig3} presents the calculated VBM and CBM of the g-C$_3$N$_4$ bilayer upon translational and rotational stacking moves with the step size of half C-N bond length. Here, the calculated electronic bands are corrected linearly by calibrating the P-6M2 bandgap to the experimental bulk value of 2.7 eV\cite{Zhang10, Wang12}, since DFT calculations are proverbial to underestimate the bandgaps of semiconductors. Accordingly, the g-C$_3$N$_4$ bilayer with P-6M2 stacking has VBM at -6.12 eV and CBM at -3.43 eV, with the vacuum level set to 0 eV. As one layer glides away from its overlapping position, such as with one step of move along the long diagonal line of the unit cell, the VBM drops sharply with 0.27 eV to -6.39 eV and its CBM lifts up with 0.08 eV to -3.35 eV, as shown in Fig. \ref{fig:fig3}a. With more translational moves, the VBM and CBM shift further up or down separately as shown in Fig. \ref{fig:fig3}a. For the rotational stacking moves, we consider initially a 60$^{\circ}$ rotation case from the P-6M2 stacking since it does not  change the unit-cell size. As shown by the inset figure in Fig. \ref{fig:fig3}, this 60$^{\circ}$-rotation stacking has the symmetry group of P-3M1. And relative to the P-6M2 bilayer, both CBM and VBM of this P-3M1 bilayer have been lifted up, respectively to -3.29 and -5.91 eV. Starting from this rotated bilayer, more translational moves result in different variation scales of VBM and CBM of the g-C$_3$N$_4$ bilayer, as shown in Fig. \ref{fig:fig3}b. Basically, stacking can be an effective parameter in tuning the positions of VBM and CBM bands of g-C$_3$N$_4$ bands,at least in scales of 0.6 and 0.3 eV respectively. That makes it a valuable parameter in optimizing the photocatalytic performance of g-C$_3$N$_4$, such as in water splitting.

\begin{figure}
      \centering
      \includegraphics[width=25em]{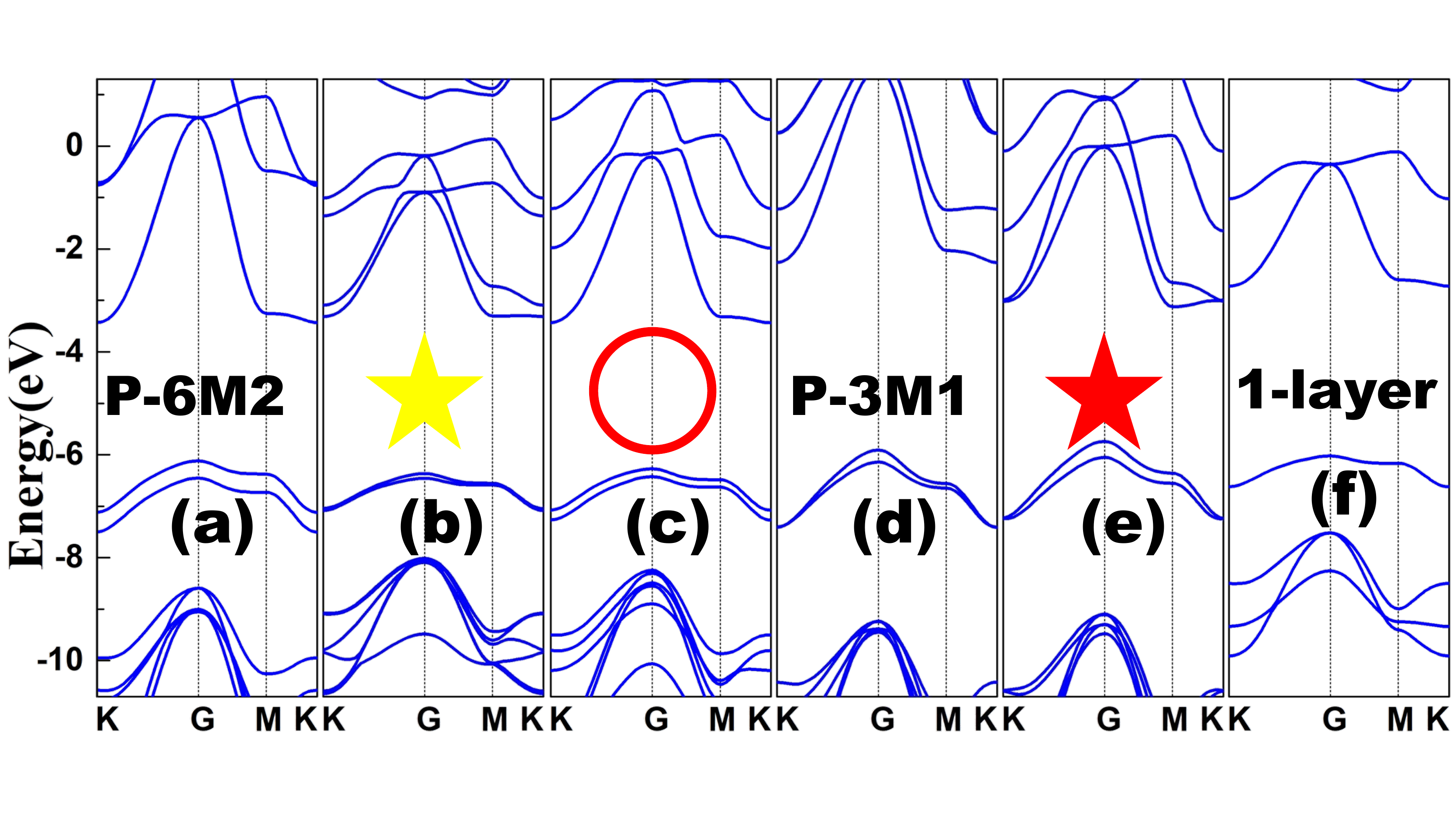}\\
      \caption{The calculated electronic band structures of g-C$_3$N$_4$ bilayers with the initial P-6M2 stacking (a), the translational stackings at the yellow star (b) and red circle (c) positions as shown in Fig. \ref{fig:fig3}a, and the initial 60$^{\circ}$-rotation stacking (d), and its translational stackings at the red star position (e) as shown in the Fig. \ref{fig:fig3}b, and the calculated electronic band structure of g-C$_3$N$_4$ single layer (f). Here the vacuum level is set to 0 eV.}\label{fig:fig4}
\end{figure}

Moreover, stacking is found also effective in tuning the electronic band profiles of g-C$_3$N$_4$ bilayers. As shown in Fig. \ref{fig:fig4}a, initially the P-6M2 bilayer has CBM at point K, and VBM at point $\Gamma$ on the top valence band (VB$_1$) followed 0.34 eV below by the second top valence band (VB$_2$). With translational moves along the long diagonal line to the yellow star stacking as labelled in Fig. \ref{fig:fig3}a, its bottom conduction band changes its dispersive profiles a little and achieves CBM at point M rather than point K, and its VB$_1$ and VB$_2$ become mixed together rather than well-separated, as shown in Fig. \ref{fig:fig4}b. As with more moves to the red circle stacking as labelled in Fig. \ref{fig:fig3}a, its CBM locates back at point K, and its VB1 and VB2 also become separated again shown in Fig. \ref{fig:fig4}c. Similarly, the bandgap nature of the initial 60$^{\circ}$-rotated bilayer changes from $\Gamma$-K (Fig. \ref{fig:fig4}d)to $\Gamma$-M (Fig. \ref{fig:fig4}e), as its layers glides from the P-3M1 stacking to the red star stacking as shown in the Fig. \ref{fig:fig3}b.
In fact, the effective inter-layer coupling between g-C$_3$N$_4$ layers originates mainly from their particular band electrons. Like graphene and h-BN\cite{Liu09}, the top valence and bottom conduction bands have the $\pi$ and $\pi$$^*$ bondings characters respectively, as shown in Fig. \ref{fig:fig1}. As touching together between the exactly overlapping layers of stacking P-6M2, the interface electrons would behave more like repulsive since they are matched with similar quantum factors. That makes their oribitals not interfering each other very much, and makes this stacking style with higher energy. Thus, the electronic bands of this P-6M2 bilayer, as shown in Fig. \ref{fig:fig4}a still keep basic profiles of the single layer as shown in Fig. \ref{fig:fig4}f, except for the well-separated VB$_1$ and VB$_2$.

\begin{figure}
      \centering
      \includegraphics[width=25em]{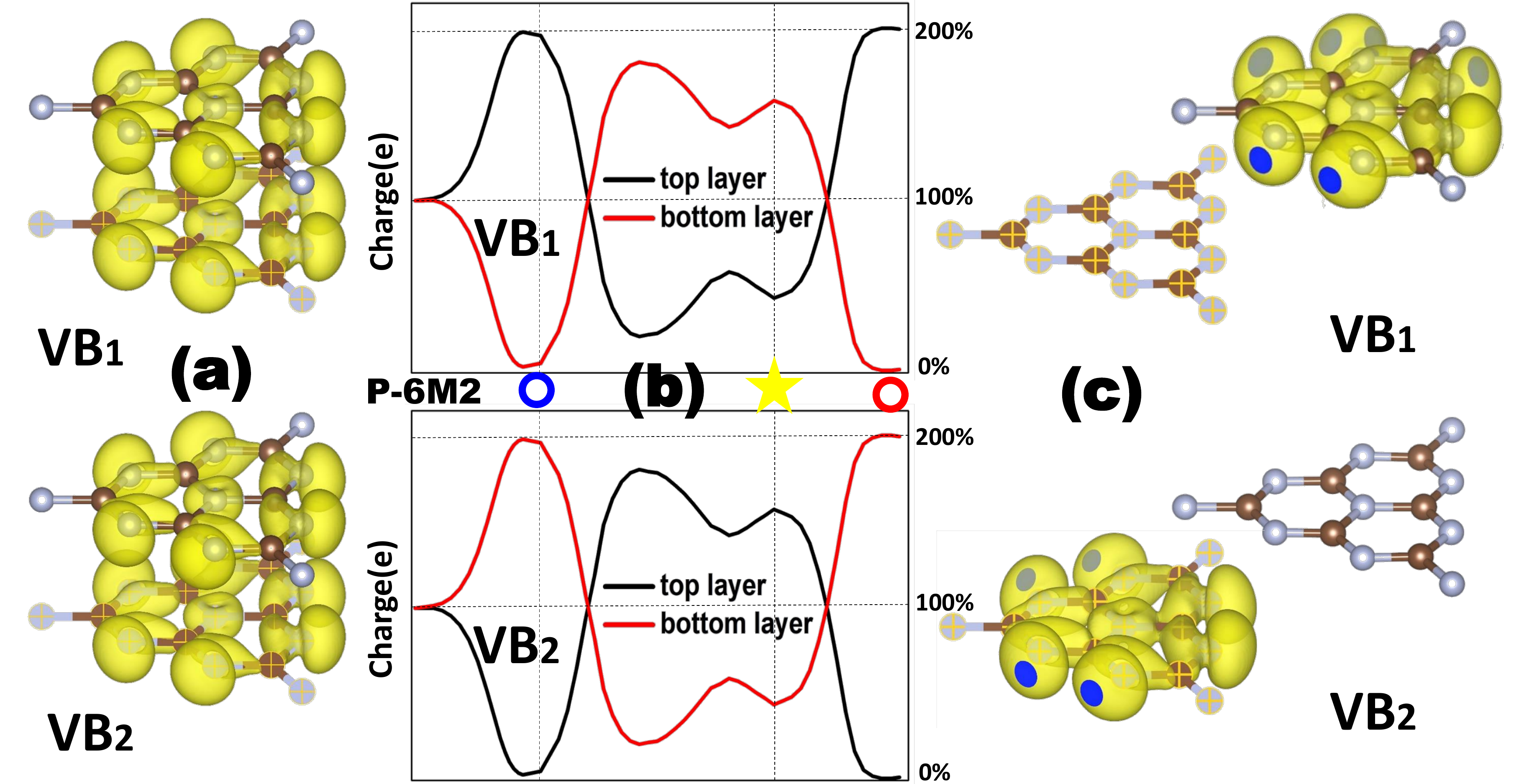}\\
      \caption{The isosurface plots of partial charge densities of VB$_1$ and VB$_2$ of the initial P-6M2 stacking (a), the dependence of integrated layer partial charges upon the translational stacking changes (b), and the isosurface plots of partial charge densities of VB$_1$ and VB$_2$ of the red-circle stacking(c).}\label{fig:fig5}
\end{figure}

Fig. \ref{fig:fig5}a plots the isosurfaces of partial charge densities of VB$_1$ and VB$_2$ of the initial P-6M2 stacking, whose  $\pi$-like orbitals distributed equally per layer and touch each other head-to-head. That is often noted as AA-stacking as in h-BN and graphite\cite{Liu03}. For the g-C$_3$N$_4$ bilayer deviated from AA-stacking, the layer electrons will couple together at interface with entangled quantum factors. That may lead to the unexpected quantum consequence of 100$\%$ spacial charge separation of band charges in van der Waals 2D layers.

Fig. \ref{fig:fig5}b plots further the charge separation dependence of g-C$_3$N$_4$ bilayers upon translational stacking moves along the long diagonal direction. In the beginning, the P-6M2 bilayer has certainly the same amount of VB$_1$ or  VB$_2$ band charges for each atomic layer. However, once the stacking starts to shift off AA-stacking, the top layer attracts more and more VB$_1$ charges from the bottom one. As the stacking approaching a blue circle stacking, the top layer can draw all VB$_1$ charges away from the bottom one. After the blue circle stacking, the bottom layer starts to draw back VB$_1$ charges from the top one, and draws the charges back and forth as stacking passing through the yellow star stacking, and eventually loses the VB$_1$ charges totally around the red circle stacking position. For the VB$_2$ band, the charge redistribution upon stack behaves reversely, i.e. initially the bottom layer grabs charges more and more from the top one in the exactly same style as the top layer does on the VB$_1$ band charges, and eventually the bottom layer grabs all VB$_2$ charges around the red circle stacking.

Such SDS charge separation is rather unusual since that has not been found in other van der Waals layers so far. Furthermore, not only g-C$_3$N$_4$ bilayers but also its multi-layers together with the bulk phase possess this SDS charge separation, as presented by their band charge plots in Support Information. The appearance of SDS charge separation in g-C$_3$N$_4$ can be ascribed to its unique triangular-mesh lattice. Each g-C$_3$N$_4$ single layer can be taken as linked together by planar C$_6$N$_8$ units, or equivalently by the triangular C$_6$N$_{10}$ molecules with shared N atoms in three corners. In that way, electrons sustains the boundary conditions to a large extent as in the isolated molecules of C$_6$N$_7$, except at the node N atoms where they have to be in tune. Therefore, as in an integrated 2D lattice of planar C$_6$N$_{10}$-linked network, the electrons will certainly have certain quantum chirality and phases with particular molecular marks, in comparison with those 2D lattices without much voids. At the interface in g-C$_3$N$_4$ layers, the inter-layer entanglement upon certain stack makes the characteristics of these particular quantum factors emerge as SDS charge separation.
This can be evidenced by that only on a few top valence bands exhibit clearly the SDS charge whereas the bottom conduction bands are free from it.

In summary, the SDS charge separation in g-C$_3$N$_4$ indeed means a new concept to separate or control charge carriers in semiconducting materials. That is intrinsically different to those conventional ways through hetero-junctions or interfaces with alien dopings or electrical fields. That makes g-C$_3$N$_4$ materials, also with proper and tunable bandgaps, have innate advantages in converting solar energy during photovoltaic or photocatalytic applications. In their photo-absorption processes, after certain decay the photo-excited charge carriers will go to the top valence band for holes and to the bottom conduction band for electrons respectively. For the ``good'' stacking, such as the red circle stacking of g-C$_3$N$_4$ bilayer, its photo-excited holes in VB1 ($\sim$150 meV above VB2) distribute only in one layer whereas its  photo-excited electrons stay in both. In principle, that SDS charge separation means the corresponding electron-hole recombination probability can be inhibited by 50$\%$. Furthermore, the SDS charge separation may add unexplored new features to the electronics of 2D materials. For example, under certain voltage VB1 or VB2 can be tuned to turn on the desired atomic-layer channel for electric current to pass through in the g-C$_3$N$_4$ bilayer or multi-layers.

\begin{acknowledgments}
L. L. acknowledges the support from the National Science Fund for Distinguished Young Scholars of China (No. 61525404).
\end{acknowledgments}

\bibliography{Ref}

\end{document}